\documentclass[twocolumn,superscriptaddress,floatfix,preprintnumbers]{revtex4}
\usepackage{graphics,amssymb,amsmath,epsfig,color}
\usepackage{graphicx}
\usepackage{amsfonts}
\usepackage{amsmath}
\usepackage{textcomp}
\usepackage{verbatim}
\usepackage{ulem}
\usepackage{epstopdf}
\usepackage{multirow}
\usepackage{eqnarray}
\usepackage{setspace} 
\usepackage{bigdelim,
			booktabs
			}	
\begin{document}

\title{Design of a spin-wave majority gate employing mode selection}

\author{S.~Klingler}
\email{klingles@rhrk.uni-kl.de}

\author{P.~Pirro}
\author{T.~Br\"acher}
\author{B.~Leven}
\author{B. Hillebrands}
\author{A.~V.~Chumak}
\affiliation{Fachbereich Physik and Landesforschungszentrum OPTIMAS, Technische Universit\"at
Kaiserslautern, 67663 Kaiserslautern, Germany}

\date{\today}

\begin{abstract}
The design of a microstructured, fully functional spin-wave majority gate is presented and studied using micromagnetic simulations. This all-magnon logic gate consists of three-input waveguides, a spin-wave combiner and an output waveguide. In order to ensure the functionality of the device, the output waveguide is designed to perform spin-wave mode selection. We demonstrate that the gate evaluates the majority of the input signals coded into the spin-wave phase. Moreover, the all-magnon data processing device is used to perform logic AND-, OR-, NAND- and NOR- operations.
\end{abstract}

\maketitle
In spintronics the degree of freedom of the spin is used to transmit information. Spin and, thus, angular momentum cannot only be transmitted by electrons, but also by magnons, the quanta of the dynamic excitations of the magnetic system - spin waves. It is possible to encode information in the phase or amplitude of such spin waves and to have it transmitted through spin-wave waveguides. Moreover, the wave properties allow for efficient data processing through the exploitation of the interference between spin waves.\cite{Schneider2008,Kostylev2005,Khitun2010,Sato2013,Xing2013, Chumak2012,Chumak2014, Bracher2013}
An important step towards the application of spin-wave devices in modern information technology is the realization of spin-wave logic gates. 
In this context, the majority gate is of special interest since it allows for the evaluation of the majority of an odd number of input signals, as given in Tab.~\ref{tab:majority_gate}. Furthermore, not only can majority operations be performed with this gate but also AND- or OR-operations, if one input (see input 3 in Tab.~\ref{tab:majority_gate}) is used as a control input. Hence, the advantage of the majority gate is its configurability and functionality.\cite{Khitun2008}
In a spin-wave majority gate the phase $\phi$ of the waves is used as an information carrier ($\phi_0$ corresponds to logic ``0'', logic ``1'' is represented by $\phi_0 + \pi$). Although the idea of such majority gates was presented earlier,\cite{Khitun2008, Cherepov2013} no practical realization suitable for the integration into magnonic circuits has thus far been proposed.

One of the main problems of a realistic spin-wave majority gate is the coexistence of different spin-wave modes with different wavelengths at a fixed frequency in the structure.\cite{Pirro2011} As a result, the output signal is given by overlaying waves of various phases and, thus, the majority function is lost. As a solution, a design which guarantees for a single-mode operation has to be used. Here, we present the design of an all-magnon majority gate and prove its functionality using numerical simulations. The width of the output waveguide has been chosen in a way such as to obtain single-mode operation. The operational characteristics of the majority gate have been studied for different phases. AND-, OR-, NAND- and NOR-operations have been demonstrated using the same majority gate device.

\begin{figure}[b]%
\begin{center}%
\scalebox{1}{\includegraphics[width=\linewidth,clip]{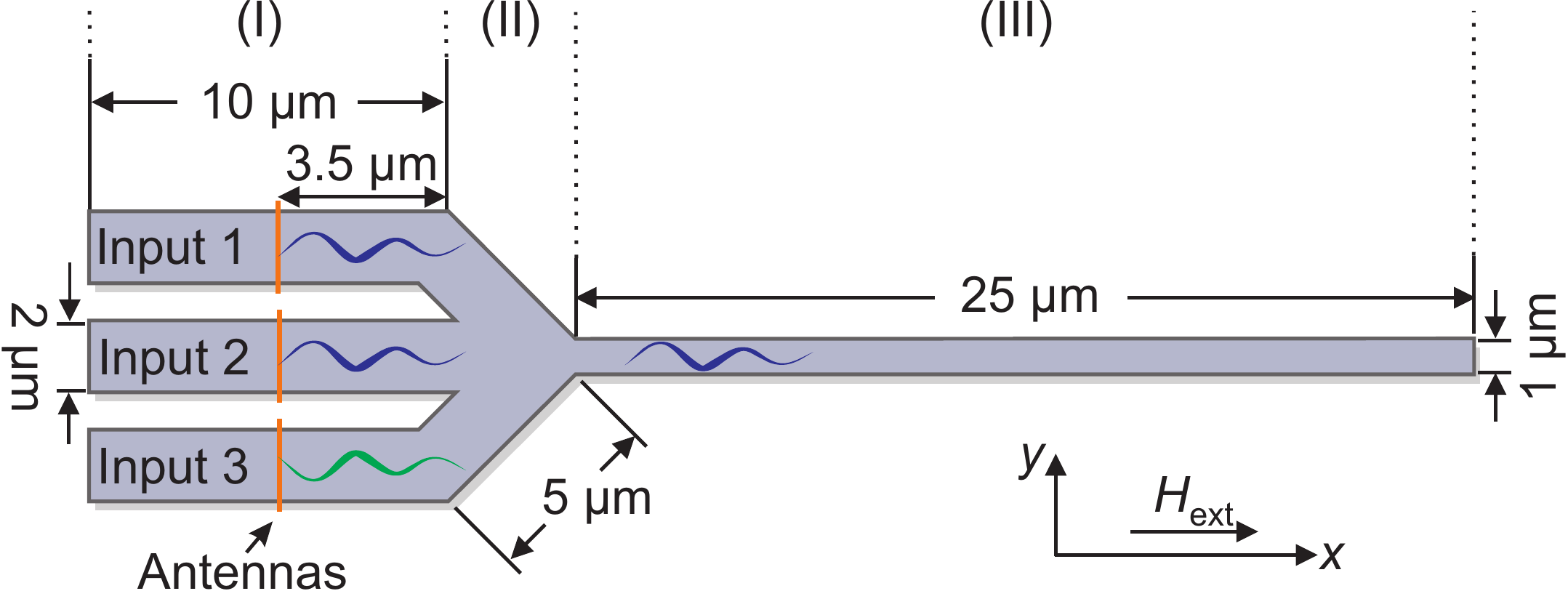}}%
\end{center}%
\caption{\label{sat_mask} 
Geometry of the majority gate. Spin waves are excited by the antennas (I), interfere in the combiner (II) and are emitted in the output waveguide (III). The phase of the spin wave in the output is determined by the phase of the incoming spin waves.}%
\end{figure}%

\begin{table}[b!]
\renewcommand{\arraystretch}{1.1}

 \begin{tabular}{ccc|clcl}
 {Input 1} & {Input 2} & Input 3 & \multirow{2}{*}{Output}&& \\
 \footnotesize{{Signal}}& \footnotesize{Signal}  &  \footnotesize{Control} & & &\\
 \cline{1-4}
 0 & 0 & 0 & 0 & \ldelim\} {4}{11pt} \multirow{4}{*}{AND} &  &\\
 1 & 0 & 0 & 0 && 	\\
 0 & 1 & 0 & 0 &&  \\
 1 & 1 & 0 & 1 && 	\\
 \cline{1-4}
 0 & 0 & 1 & 0 &  \ldelim\} {4}{11pt} \multirow{4}{*}{OR} & & \\
 1 & 0 & 1 & 1 && 	\\
 0 & 1 & 1 & 1 &&   \\
 1 & 1 & 1 & 1 && 	\\
 \end{tabular}
  \caption{Truth table of the majority operation. The majority of the input signals passes through the gate. In addition, input 3 can be seen as a control gate which switches the function of the gate between AND ($\text{Input 3}=0$) and OR ($\text{Input 3}=1$) operation. NAND- and NOR-operations can be performed through shifting the phase of the output signal by $\pi$.}
 \label{tab:majority_gate}
 \end{table}

For the simulations, the material parameters of 100\,nm-thick Yttrium-Iron-Garnet (YIG) are used: a saturation magnetization\cite{Algra1982} of $M_\mathrm{s}=140$\,kA/m, an exchange constant\cite{Anderson1964} of $A=3.5$\,pJ/m and a Gilbert damping\cite{Pirro2014} of $\alpha=5\cdot 10^{-4}$ which are realistic values for YIG films of nanometer thickness. The use of YIG is motivated by its large intrinsic spin-wave propagation distance which is larger than the size of the microstructures, \cite{Serga2010,Pirro2014,Hahn2014} allowing for the construction of complex magnonic networks.\cite{Chumak2014}

\begin{figure}[t]%
\begin{center}%
\scalebox{1}{\includegraphics[width=\linewidth,clip]{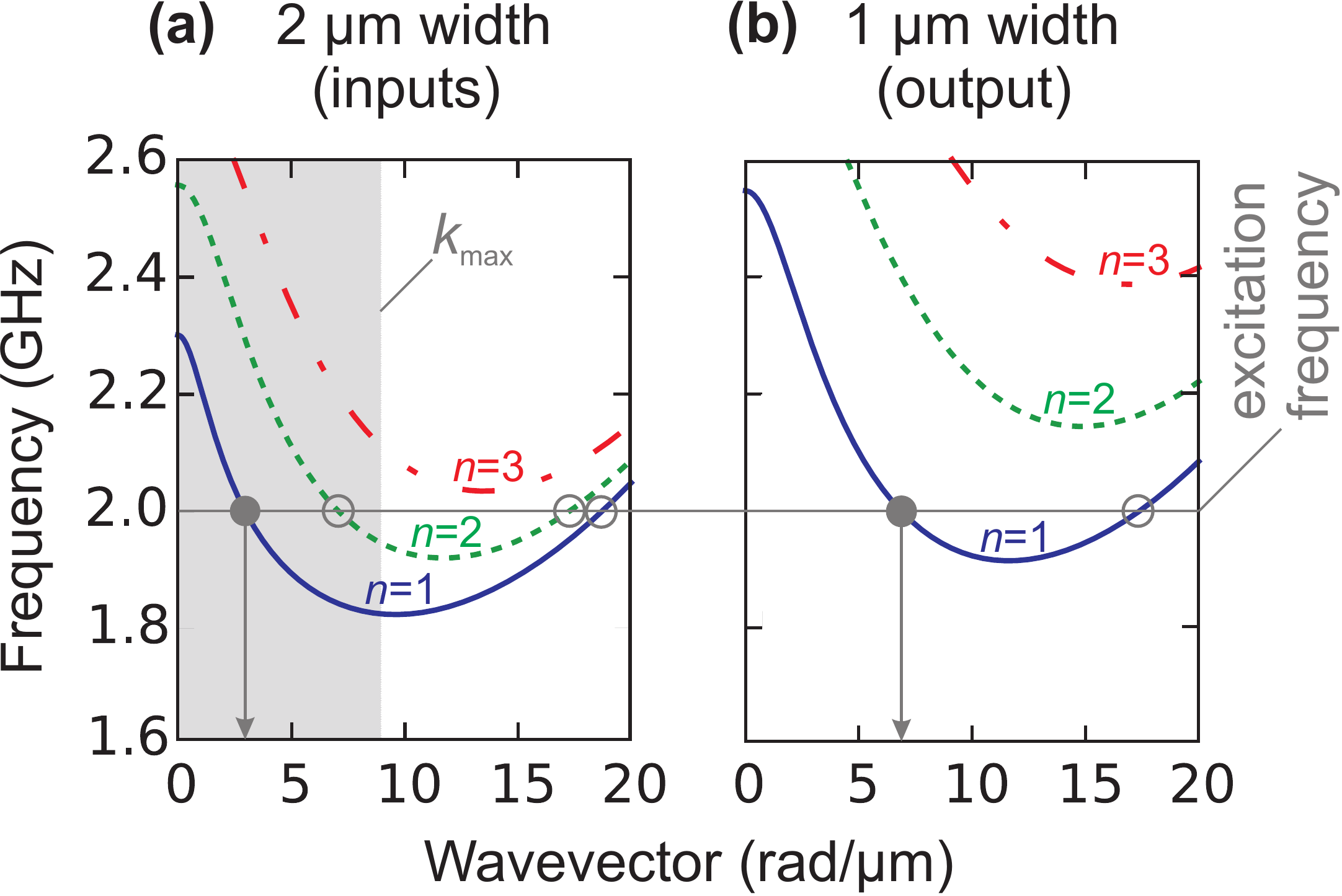}}%
\end{center}%
\caption{\label{disp_relation} 
Dispersion characteristics of a 2\,$\mu$m and a 1\,$\mu$m wide waveguide in backward-volume geometry and an external magnetic field of 24\,mT. Dynamic strayfields due to the finite size of the microstructures were considered using an effective stripe width.\cite{Guslienko2002} (a) The first and the second width mode can exist in the waveguide. The regime with an efficient spin-wave excitation is shaded in grey. (b) At a frequency of 2\,GHz only the first width mode can exist in the waveguide.}%
\end{figure}%
	
The design of the investigated majority gate is shown in Fig. \ref{sat_mask}. It can be divided into three areas: The first area (I) consists of three input waveguides where spin waves can be excited with different phases. The second area (II) is the spin-wave combiner where the outer waveguides are merged into the central waveguide at an angle of 45$^\circ$. The area (III) is defined by the output waveguide where spin waves propagate with the same phase as the majority of the incoming spin waves. To ensure the possibility of a practical experimental realization of the gate, all dimensions are chosen to be in the micrometer range.\cite{Chumak2014,Pirro2014,Hahn2014}

The input waveguides (I) have a width of $w=2\,\mu$m and a length of 10\,$\mu$m. In this area, the spin-wave excitation by micropstrip antennas crossing the waveguides orthogonally is modeled. 
The magnetization and the wavevector point in the direction parallel to the long axes of the waveguide. For this configuration the dispersion relations\cite{Kalinikos1986} of different width modes $n$, where $n$ refers to the number of anti-nodes across the width of the waveguide, are shown in Fig. \ref{disp_relation}(a). For low values of $k$, where the spin waves are known as backward-volume magnetostatic waves, the frequency decreases with increasing wave vector due to dipolar interactions. Thus, spin waves with low $k$ are in the dipolar spin wave regime.
For large values of $k$ the frequency increases with increasing wave vector due to exchange interaction, and the spin waves are in the exchange regime. 
The excitation efficiency of the microstrip antenna decreases with increasing wave vector.\cite{Demidov2009} This allows for the excitation of the dipolar spin waves at a frequency of 2\,GHz with sufficient efficiency while the excitation of the exchange waves can be neglected (marked with filled and unfilled dots, respectively). The even numbered width modes cannot be excited with an homogeneous antenna field.\cite{Kittel1958} Thus, only the first width mode was employed in the transmitting and processing of information in the investigated majority gate.
In the spin-wave combiner (part II), the length of the bends was chosen to be a multiple of the spin-wave wavelength, in order that changes in the phased due to the propagation through the bend were prevented.
However, scattering into higher width modes can occur in the combiner since in the bend the transitional symmetry is broken.\cite{Clausen2011,Noether1918}
To avoid these higher-order width modes in the output waveguide (III), its width is chosen to be 1\,$\mu$m. As can be seen in Fig. \ref{disp_relation}(b), the minima of the dispersion relations are shifted to higher frequency values for the 1\,$\mu$m wide stripe due to the smaller quantization width. This can be used to shift the dispersion relation of higher order modes beyond the excitation frequency of 2\,GHz. Thus, for this frequency, higher width modes are prohibited in the output waveguide. Additionally, the wavevector of the spin wave is increased in the 1\,$\mu$m-wide stripe to $k=6.3\,\mu$m$^{-1}$ (cf. gray vertical arrows in Fig. \ref{disp_relation}).

To investigate the spin-wave dynamics in the proposed structure (shown in Fig. \ref{sat_mask}), numerical simulations using MuMax2, an approach of parallel computation in micromagnetic simulations using graphics cards (GPUs), \cite{Vansteenkiste2011} were performed. This allows for the simulation of the spin-dynamics in a micron-sized area with a resolution of a few nanometers.\cite{Pirro2011} To avoid spin-wave reflections at the end of the waveguide, the damping is increased continuously by a factor of 300 over the last 4\,$\mu$m in $x$-direction. A static field of 24\,mT is applied in $x$-direction, and the spin waves propagate parallel to the magnetization. Furthermore, the simulation includes dipol-dipol- and exchange-interaction based on the aforementioned values. The size of the entire simulated area in Fig. \ref{sat_mask} is $8.07\times38.54\times0.1$\,$\mu$m$^3$ and was divided into $768\times2048\times1$ cells. Thus, the cells have a cuboidal shape with a size of approximately $10\times19\times100$\,nm$^3$. In the lateral directions the cell size is in the order of the exchange length ($\lambda_\mathrm{ex}=18$\,nm),\cite{Abo2013} which is smaller than the wavelength of the spin-wave. The approximation of a uniform magnetization profile across the thickness is justified since no perpendicular standing spin-wave modes can be excited at the excitation frequency of 2\,GHz. To model the spin-wave excitation, an alternating magnetic field which is produced by a microstrip antenna with a width of 700\,nm and a height of 400\,nm was assumed. 
For this, Biot-Savart's Law is used for a rectangular AC-current distribution with a current amplitude of 0.1\,mA and a frequency of 2\,GHz. 
The dynamic excitation was simulated for 100\,ns to ensure that the amplitudes of the spin waves did not increase further and are in balance with the damping. Then, the magnetization configuration was saved with a time resolution of 4\,ps.

\begin{figure}[t!]%
\begin{center}%
\scalebox{1}{\includegraphics[width=\linewidth,clip]{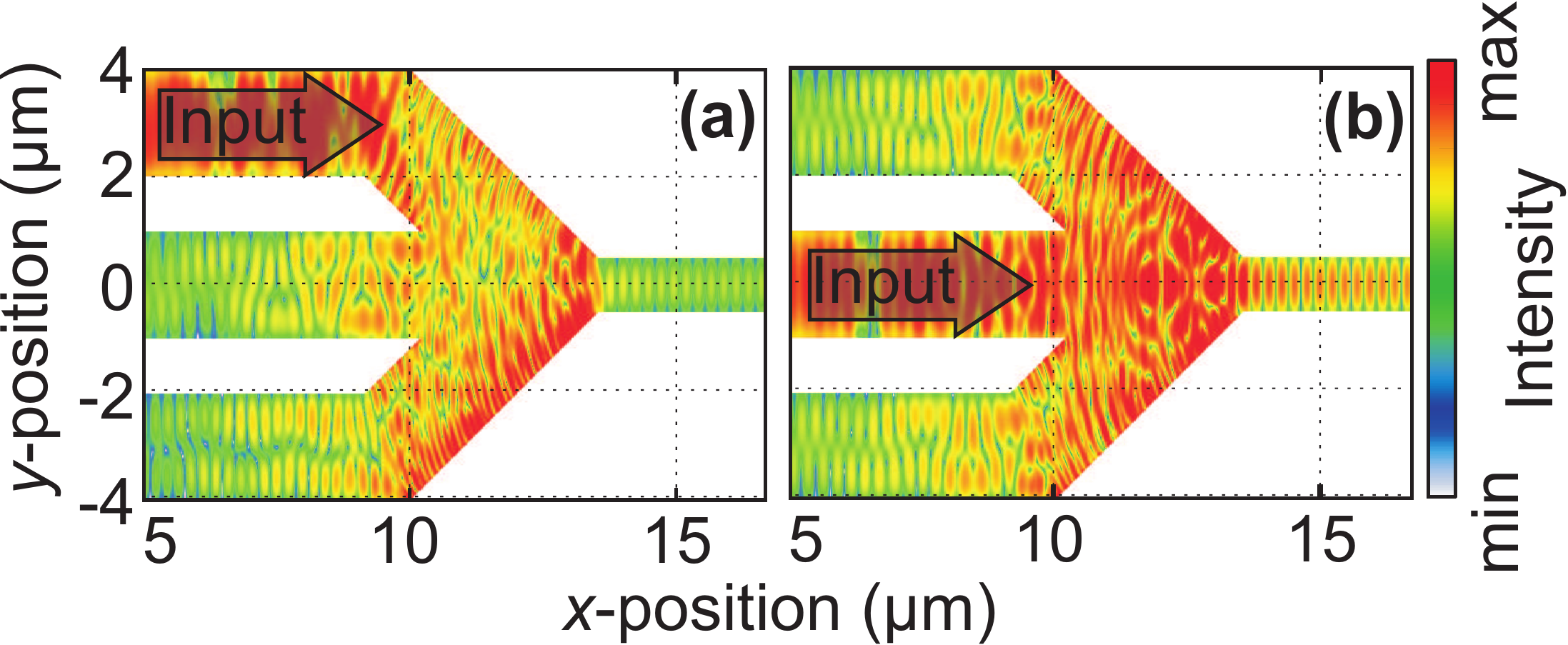}}%
\end{center}%
\caption{\label{single_arm} 
Time averaged intensity distribution for a single arm excitation. The color code is in a logarithmic scale. (a) The upper arm of the majority gate is excited, and a strong intensity flow into the lower arm is observed. The spin wave intensity is only partially emitted into the output arm. (b) The central input arm is excited. The emission of intensity is symmetric into the other input arms. The intensity distribution in the output arm is much higher in comparison to the upper arm excitation.}%
\end{figure}%

At first, the spin-wave dynamics in the combiner were investigated if only one arm is excited. By this, information about the intensity and the mode distribution in the structure can be obtained. In Fig. \ref{single_arm}(a), the time averaged intensity is shown when only the upper arm is excited. If the spin wave is excited in the lower arm, the intensity distribution is symmetric to the $y=0\,\mu$m-axis (not shown). The spin waves in the bend have a wavevector component in $y$-direction, which is perpendicular to the static magnetization. The group velocity component in the $y$-direction is larger than that of the component parallel to the magnetization. Thus, the intensity flow into the lower arm is preferred in comparison to the output waveguide. Furthermore, the second width mode is excited in the other input arms due to the scattering processes in the combiner. In Fig. \ref{single_arm}(b), the intensity distribution is shown if the spin wave is excited in the central arm. The intensity flow into the output waveguide is increased in comparison to the situation in Fig.~\ref{single_arm}(a), since the spin waves do not have to bend around a corner to reach the output waveguide. Thus, the amplitude of the spin-wave excitation in the central arm has to be attenuated by a factor of 0.2 in order to equalize the amplitudes of the output signals. In both cases, higher modes are excited in the combiner due to its large quantization width. However, it has to be emphasized that only the first width modes (dipolar and exchange) exist in the output waveguide which determine the shape of the interference patterns visible in the output waveguide.

\begin{figure}[t]%
\begin{center}%
\scalebox{1}{\includegraphics[width=\linewidth,clip]{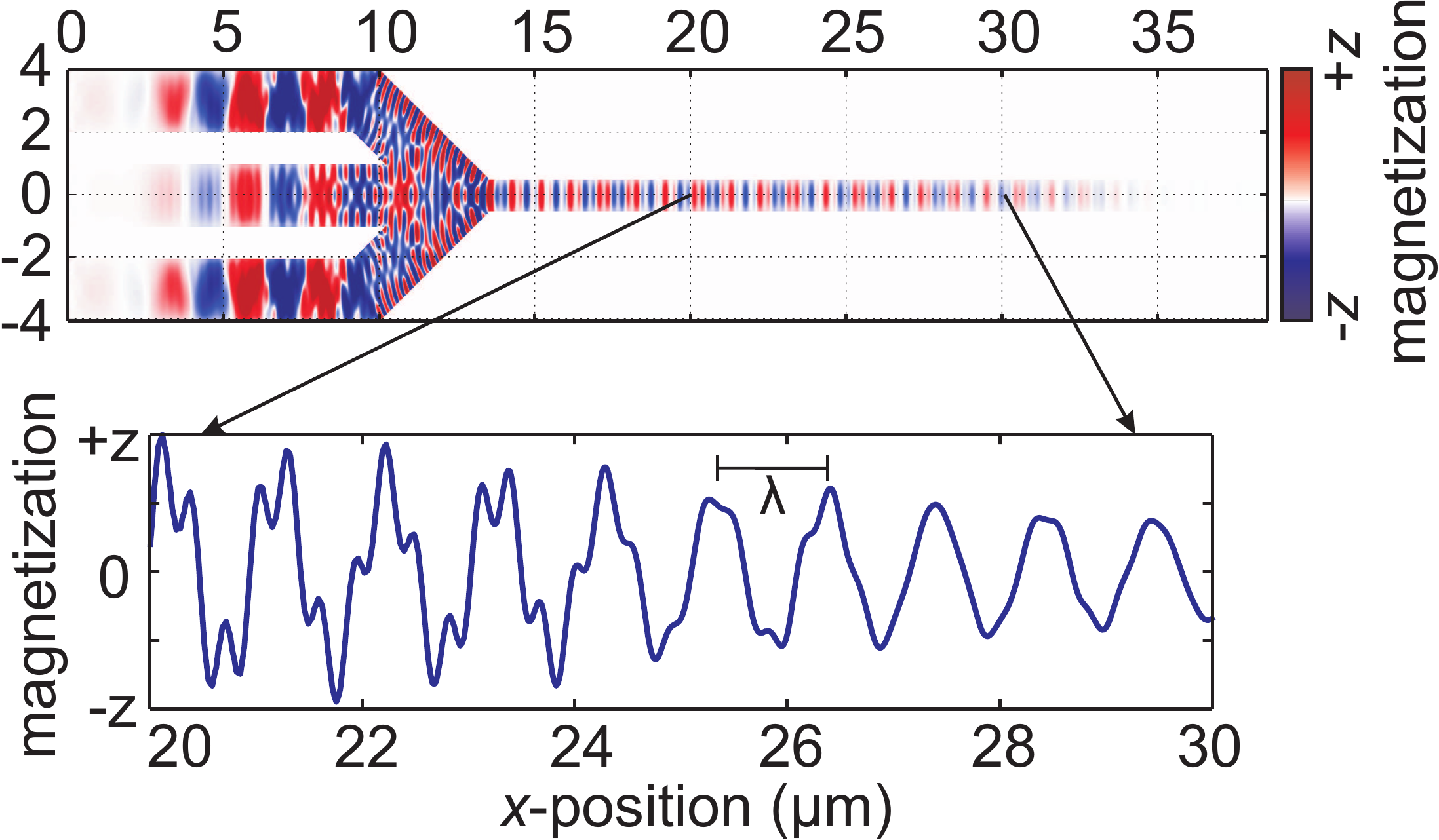}}%
\end{center}%
\caption{\label{z_mag} 
Top: $z$-magnetization at the last time step of the 0-0-0-simulation (the spin waves in all inputs are excited without additional phase shift). The color code in the upper graph displays the deviations in the magnetization distribution. Bottom: Extracted $m_z$ in area (III) shows a mode interference between the dipolar and the exchange spin wave in the output waveguide.}%
\end{figure}

The previous section explored the spin-wave intensity contributions in the combiner. Now, the operational characteristics of the majority gate are discussed for the cases where all inputs are excited. For this, the phase information of the combined spin waves in the output waveguide needs to be extracted. This is achieved by comparing the $z$-magnetization $m_z$ at a fixed timestep for every input combination. A snapshot of $m_z$ from the 0-0-0-simulation (i.e., in the first, second and third input spin waves are excited without additional phase shift) is shown in the upper graph in Fig. \ref{z_mag}. In the lower graph, $m_z$ at the center of the output waveguide is extracted. It can be seen that the observed signal is a superposition of two waves with different wavelength. The large wavelength ($\lambda \approx1$\,$\mu$m and $k \approx6.3$\,$\mu$m$^{-1}$) can be identified with the dipolar wave, the small wavelength can be attributed to the exchange wave in the dispersion relation in Fig. \ref{disp_relation}(b). The influence of the exchange wave decreases with an increasing distance $x$ from the combiner since its group velocity is smaller ($v_\mathrm{gr}^\mathrm{ex}\approx0.18\,\frac{\,\mu \text{m}}{\text{ns}}$) than the group velocity of the dipolar wave ($v_\mathrm{gr}^\mathrm{dipolar}\approx0.28\,\frac{\,\mu \text{m}}{\text{ns}}$). Thus, the exchange wave has a shorter decay length and its influence can be neglected for large values of $x$.

\begin{figure*}[t!]%
\begin{center}%
\scalebox{1}{\includegraphics[width=0.95\linewidth,clip]{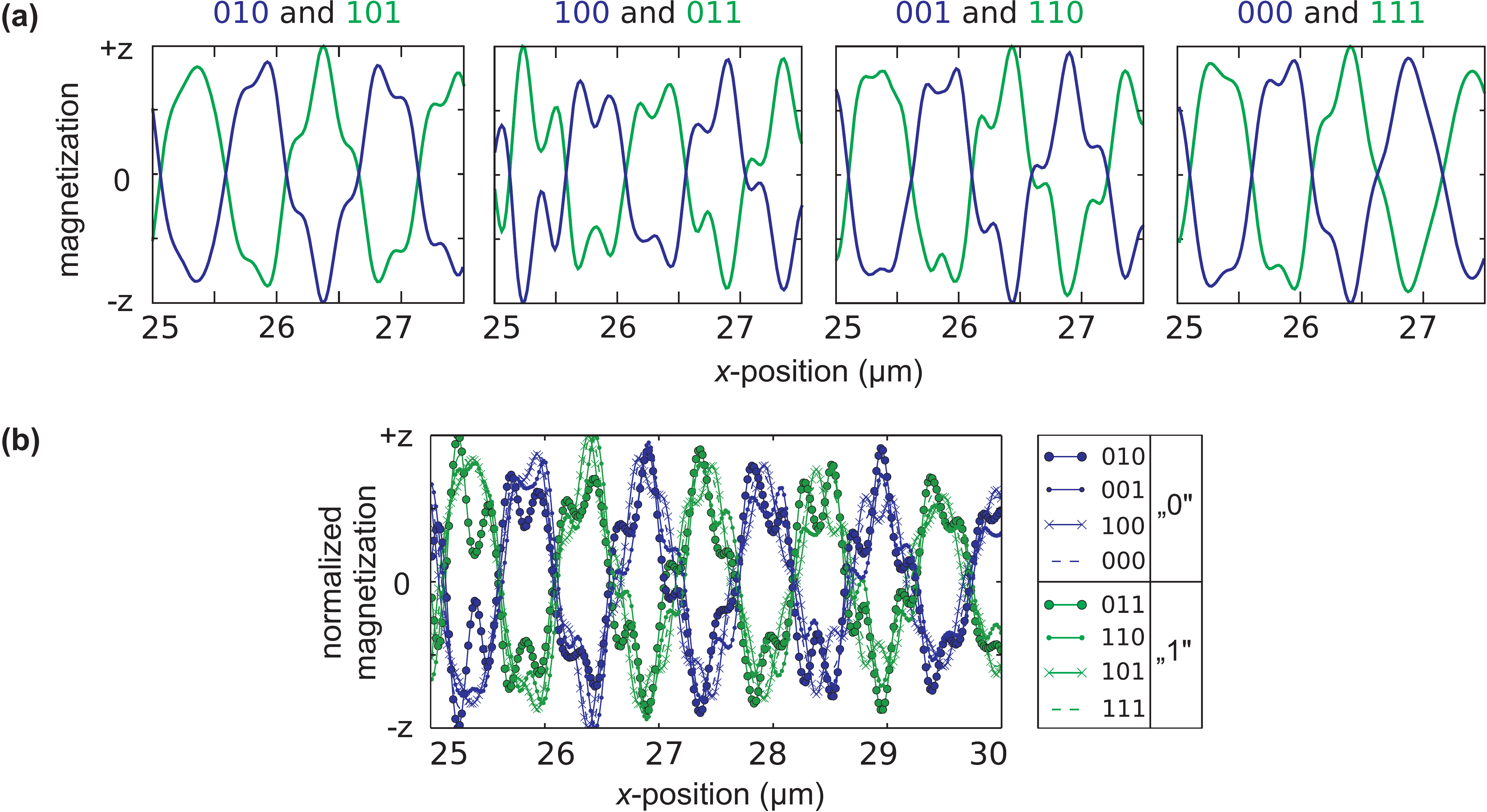}}%
\end{center}%
\caption{\label{full_majority} 
(a) Normalized magnetization distribution of different input combinations. The shift of the majority phase in the input waveguides leads to a phase shift of $\pi$ in the output waveguide. (b) All input combinations are shown in one graph. Input combinations with the same majority phase are in phase. Input combinations with different majority phase are out of phase.}%
\end{figure*}%

To compare the phase information of every possible input combination, the $z$-magnetization at the same spatial position and at the same point in time of the simulations is extracted for each combination. 
In Fig. \ref{full_majority}(a) pairs of the extracted $z$-magnetization are shown, where for each pair each input has been shifted to its opposite phase, e.g., 0-1-0 to 1-0-1 in the left panel. For these input combinations, the logic majority value has been shown to be different without a change of the relative phases between the inputs, e.g., in the aforesaid example inputs 1 and 3 are in phase for both combinations. As one can see in Fig. \ref{full_majority}(a), the spatial magnetization distributions in the output waveguide (III) are exactly out of phase and symmetric with respect to the $m_z=0$-axis due to the change of the majority value. As mentioned before, the signal is a superposition of both the dipolar and exchange waves.
If the outer input waves are excited in phase (as in the right hand panel), the phase-sensitive coupling into the first dipolar mode is stronger due to the symmetric spin-wave excitation at the input of the output waveguide. Thus, the dipolar wave is excited more effectively and the occurrence and influence of the exchange mode can almost be neglected. Furthermore, the output amplitude depends on the exact input combination. However, since the information is transmitted and processed in the phase of the spin wave, the output amplitude is of minor interest.

To proof that the shown waveguide arrangement works as a majority gate, the output phases of all input combinations are compared in Fig. \ref{full_majority}(b). It is clearly visible that all excitation combinations with a majority phase of 0 (blue) are in phase, and simultaneously in anti-phase to all combinations with a majority phase of $\pi$ (green). With this, the majority operation from Tab. \ref{tab:majority_gate} is reproduced and it is shown that the phase of the spin wave in the output is always defined by the phase of the majority of the input waves.

In addition, it has to be highlighted that the investigated all-magnon majority gate also allows for a full set of spin-wave logic operations (AND, OR, NAND and NOR). For this purpose, input 3 has exemplarily been chosen as a control input (see Tab. \ref{tab:majority_gate}). For a fixed readout position, e.g., $x=29$\,$\mu$m, AND-operations are performed when the spin waves at input 3 are excited without an additional phaseshift, i.e., the logic value of input 3 is ``0''. By switching the phase in input 3 to $\pi$, i.e., the logic value of input 3 becomes ``1'', OR-operations are performed at the fixed readout position. By shifting the readout position by a half of the wavelength of the dipolar wave, e.g., from $x=29$\,$\mu$m to $x=29.5$\,$\mu$m, the phase of the readout signal is shifted by $\pi$ and all logic output values are negated. Thus, NAND- and NOR-operations are simultaneously performed dependent on the control input. This is a general benefit of coding data into the spin-wave phase.\cite{Khitun2008}

In summary, a fully operational spin-wave majority gate has been presented, where data transmission and processing are realized with spin waves.
By choosing the proper width of the output waveguide, it is possible to select the first width mode from the combiner, since the dispersion relations of the higher modes can be shifted above the excitation frequency with decreasing waveguide width.
It was demonstrated, that the majority of the logic values of the input signals determines the logic value of the output signal with the proposed geometry. It was shown, that all excitation combinations with a phase majority of 0 are in phase and simultaneously in anti-phase to all combinations with a phase majority of $\pi$. In addition, logic AND- and OR-operations can be performed by choosing a fixed readout positions in the output waveguide, whereas NAND- and NOR-operations can be performed by shifting the readout position by a half of the spin-wave wavelength. It has to be emphasized, that the data processing in the proposed majority gate occurs fully in the magnonic system. With this, the output signal can be directly used in combination with an all-magnon transistor \cite{Chumak2014} and various spin-wave devices.\cite{Demidov2009a, Bracher2013, Vogt2014}

This research has been supported by the EU-FET grant InSpin 612759.

\end{document}